\documentclass[aps,physrev,reprint,noeprint]{revtex4-1} 
\usepackage{bm, braket,amsmath,mathtools,comment}
\usepackage{graphicx,ulem}
\usepackage{hyperref}
\usepackage{color}

\begin{document}
\title{Labeling Eigenstates of Qubit-Cavity Systems Based on the Continuity of Qubit Occupancy: Detecting Resonances to Higher Excited Qubit States}
\author{Shimpei Goto}
\email[]{goto.las@tmd.ac.jp}
\author{Kazuki Koshino}
\email[]{kazuki.koshino@osamember.org}
\affiliation{Institute for Liberal Arts, Institute of Science Tokyo, Ichikawa, Chiba 272--0827, Japan}
\date{\today}
\begin{abstract}
    We propose a new method for labeling the eigenstates of qubit-cavity systems based on the continuity of the qubit occupancy.
    The labeled eigenstates give a rough estimate of the evolution of a quantum state under cavity driving.
    The photon-number dependence of the resonant cavity frequency can be estimated from the labeled eigenenergies, and resonances to higher excited qubit states are visible in the dependence.
    Our proposed method can be applied to a broader situation compared to an existing method.
    With the proposed method, we investigate the offset-charge dependence of the resonances to higher excited states that can induce leakage effects from the computational basis.
    The results imply that the leakage can occur with only around ten photons.
\end{abstract}
\maketitle
\section{Introduction\label{sec:introduction}}
A system composed of a qubit and a cavity is one of the fundamental setups to study the quantum light-matter interaction, as demonstrated in the seminal Jaynes-Cummings model~\cite{jaynes_comparison_1963}.
Moreover, the system is an essential building block for superconducting quantum computation since the readout of the qubit state can be performed with the setup by the dispersive readout~\cite{blais_cavity_2004,wallraff_strong_2004,vijay_observation_2011}.
The dispersive readout is a ubiquitous method in the sense that this readout scheme can be applied to any type of qubit.
The importance of qubit-cavity systems is still growing with the development of quantum information technologies.

The qubit used in the superconducting circuits has some complexities that are absent in the Jaynes-Cummings model.
The number of states in the qubit is unbounded in general~\cite{bouchiat_quantum_1998,nakamura_coherent_1999,orlando_superconducting_1999,yan_flux_2016,martinis_rabi_2002,koch_charge-insensitive_2007,vool_introduction_2017,blais_circuit_2021,manucharyan_fluxonium_2009,kalashnikov_bifluxon_2020,patel_d-mon_2024,brosco_superconducting_2024}.
The fluctuation of charged impurities causes the temporal variation of qubit frequency and leads to the decoherence of the qubit~\cite{nakamura_coherent_1999,metcalfe_measuring_2007,blais_circuit_2021}.
The counter-rotating terms neglected so often~\cite{jaynes_comparison_1963,tavis_exact_1968} can induce the transition to a higher excited qubit state~\cite{sank_measurement-induced_2016}.
These complexities should be considered when one studies qubit-cavity systems for practical applications.

Besides, the dispersive readout inherits these complexities.
The working principle of the dispersive readout is a qubit-state-dependent frequency shift to the cavity~\cite{blais_cavity_2004,blais_quantum-information_2007,vijay_observation_2011,blais_circuit_2021}.
Higher excited qubit states also affect the frequency shift, and the weight of such excited states depends on the cavity photon number.
Consequently, the cavity resonant frequency also depends on the cavity photon number~\cite{shillito_dynamics_2022-1,goto_efficient_2023,khezri_measurement-induced_2023}.
It is required to evaluate the photon-number dependence of the cavity frequency, especially for operating the readout with high photon numbers.
The transition to a higher excited state leads to the leakage of a quantum state from the computational space~\cite{sank_measurement-induced_2016,lescanne_escape_2019,verney_structural_2019,khezri_measurement-induced_2023}.
Thus, it is also necessary to estimate locations where such transitions occur.

In this paper, we propose a new approach to label the eigenstates of qubit-cavity systems, which can be applied to a broader situation compared to an existing method.
The photon-number dependence of the cavity frequency can be estimated by labeling the eigenstates with a qubit state and a photon number.
A resonance to a higher excited state, which can lead to the transition to the excited state, is visible in the photon-number dependence of the cavity frequency~\cite{shillito_dynamics_2022-1,goto_efficient_2023}.

There already exist some approaches to label the eigenstates~\cite{boissonneault_improved_2010,khezri_measuring_2016,shillito_dynamics_2022-1,xiao_diagrammatic_2023}.
Floquet modes in driven systems are labeled~\cite{xiao_diagrammatic_2023} as well as the eigenstates of undriven systems.
Some of these approaches utilize the conservation of the total number of excitations~\cite{boissonneault_improved_2010,khezri_measuring_2016} and cannot be applied to situations where the counter-rotating terms are not negligible.
For the method applicable to systems with counter-rotating terms~\cite{shillito_dynamics_2022-1}, we find instances where the existing method gives discontinuous photon-number dependence of qubit occupancy.
In such instances, the obtained labeling follows a resonant transition, and one cannot estimate the cavity frequency for a case where the transition is avoided by quickly passing the resonant point.
Thus, we develop and propose a new approach that estimates continuous dependence even in such cases.
Comparing the photon-number dependence obtained by the existing and the proposed approaches to the evolution under strong photon driving enough to avoid the resonant transition, we observe that the evolution of the qubit occupancy roughly follows the photon-number dependence given by the proposed method.

With the proposed method, we also investigate the offset-charge dependence of resonances to higher qubit excited states in the transmon-cavity system.
Contributions from charged impurities are encapsulated in the offset charge.
We observe that the locations of the resonances strongly depend on the offset charge and that only around ten photons can induce a transition leading to leakage from the computational space.

The rest of the paper is organized as follows: In Sect.~\ref{sec:method}, we introduce labeling methods used in this paper.
The proposed approach is given in this section.
In Sect.~\ref{sec:comparison}, we compare the existing and the proposed labeling methods.
The behavior of the proposed method can be controlled by the energy window.
We discuss how the choice of the energy window affects the labeling in Sect.~\ref{sec:energy_window}.
In Sect.~\ref{sec:result}, the offset-charge dependence of resonances to higher excited qubit states is investigated with the proposed method.
The summary is given in Sect.~\ref{sec:summary}.

\section{Labeling methods\label{sec:method}}
The purpose of this paper is to propose a new method to label the eigenstates of qubit-cavity systems.
The Hamiltonian of the qubit-cavity system is given by
\begin{align}
    \hat{H} = \sum^\infty_{i = 0}\hbar \omega_i \ket{i}_q \bra{i}_q + \hbar \omega_c \hat{c}^\dagger \hat{c} + \hat{H}_\mathrm{int}.
\end{align}
Here, \(\ket{i}_q\) is the \(i\)--th excited eigenstate of the qubit whose eigenenergy is \(\hbar \omega_i\) (\(\ket{0}_q\) corresponds to the ground state), \(\omega_c\) is the resonant frequency of the cavity, \(\hat{c}\) (\(\hat{c}^\dagger \)) denotes the bosonic annihilation (creation) operator for the cavity, and \(\hat{H}_\mathrm{int}\) denotes the interaction between the qubit and the cavity.
The ground (first excited) state of the qubit \(\ket{0}_q\) (\(\ket{1}_q\)) is also denoted by \(\ket{g}_q\) (\(\ket{e}_q\)).
We assume the validity of the perturbative treatment of the interaction term \(\hat{H}_\mathrm{int}\) for \(\ket{g}_q \otimes \ket{0}_c\) (\(\ket{e}_q \otimes \ket{0}_c\)) so that one can uniquely find the state \(\ket{\overline{g}, \overline{0}}\) (\(\ket{\overline{e}, \overline{0}}\)) from the eigenstates of \(\hat{H}\) which has the largest overlap with \(\ket{g}_q \otimes \ket{0}_c\) (\(\ket{e}_q \otimes \ket{0}_c\)).
Here, \(\ket{i}_c\) is the Fock state of the cavity which satisfies \(\hat{c}^\dagger \hat{c} \ket{i}_c = i \ket{i}_c\).

We discuss how to find a labeled state for a qubit state \(\ket{p}_q\) and a Fock state \(\ket{n}_c\), \(\ket{\overline{p}, \overline{n}}\), from the eigenstates of \(\hat{H}\).
Starting from an initial state \(\ket{\overline{p}, \overline{0}}\), the system is expected to follow the labeled states \(\ket{\overline{p}, \overline{n}}\) under adiabatic injection of photons into the cavity~\cite{boissonneault_improved_2010,goto_efficient_2023,shillito_dynamics_2022-1}. 
The eigenenergy of the labeled eigenstate \(\ket{\overline{p}, \overline{n}}\) is denoted by \(\varepsilon_{p,n}\).
With the labeled eigenenergies, one can regard \((\varepsilon_{p, n+1} - \varepsilon_{p, n})/\hbar \) as the effective cavity frequency when the qubit is initially \(\ket{p}_q\) and the cavity photon number is \(n\).
The photon-number dependence of the cavity frequency is useful information for designing the parameters of the dispersive readout~\cite{shillito_dynamics_2022-1,goto_efficient_2023}.
Besides, a resonance to a higher excited qubit state becomes visible in the photon-number dependence of the cavity frequency.

The simplest approach to find the labeled state \(\ket{\overline{p}, \overline{n}}\) is the overlap approach: an eigenstate which has the largest overlap with a product state \(\ket{p}_q \otimes \ket{n}_c\) is \(\ket{\overline{p}, \overline{n}}\).
This method does not work in a large \(n\) region, as we will present in Sect.~\ref{sec:comparison}.
The failure of the overlap approach means that \(\ket{\overline{p}, \overline{n}}\) cannot be represented as \(\sqrt{1-|\epsilon|^2}\ket{p}_q \otimes \ket{n}_c + \epsilon \ket{\phi}\) for the large \(n\) region, where \(\epsilon \) is some small constant and \(\ket{\phi}\) is a state orthogonal to \(\ket{p}_q \otimes \ket{n}_c\).

When the Hamiltonian \(\hat{H}\) preserves the total excitation number, i.e., \([\hat{H}, \hat{N}_q+\hat{c}^\dagger \hat{c}] = 0\), the method based on block diagonal structure is available~\cite{boissonneault_improved_2010,khezri_measuring_2016}.
Here, \(\hat{N}_q\) is the qubit occupation number operator defined by
\begin{align}
    \hat{N}_q = \sum^\infty_{i=1} i \ket{i}_q \bra{i}_q.
\end{align}
In such cases, the Hamiltonian has a block diagonal structure with respect to the total number of excitations \(\braket{\hat{N}_q + \hat{c}^\dagger \hat{c}}\).
When the eigenstate \(\ket{\lambda}\) belongs to the block corresponds to the total excitation number \(M\) and the cavity photon number \(\braket{\lambda|\hat{c}^\dagger \hat{c}|\lambda}\) is close to \(n\), \(\ket{\lambda}\) should be \(\ket{\overline{M-n}, \overline{n}}\).
This method cannot be applied when \(\hat{H}\) does not preserve the total excitation number, and the resonance resulting from non-preserving terms has been reported in superconducting qubit systems~\cite{sank_measurement-induced_2016}.
Thus, a method applicable to non-preserving systems is preferred.

A recursive approach proposed in Ref.~\cite{shillito_dynamics_2022-1} can be applied to non-preserving systems.
Starting from \(\ket{\overline{p}, \overline{0}}\) which can be obtained by the overlap approach, a state \(\ket{\overline{p}, \overline{n}}\) is obtained from \(\ket{\overline{p}, \overline{n-1}}\) recursively.
In the recursive approach, a state \(\hat{c}^\dagger \ket{\overline{p}, \overline{n-1}}\) is used as a candidate for the next labeled state \(\ket{\overline{p}, \overline{n}}\).
Therefore, \(\ket{\overline{p}, \overline{n}}\) is determined as an eigenstate which has the largest overlap with the state \(\hat{c}^\dagger\ket{\overline{p}, \overline{n-1}}\).
This labeling method is used to analyze the readout dynamics of the transmon qubit with high-power input light, and the effective cavity frequency obtained by this method reflects the resonance coming from non-preserving terms~\cite{shillito_dynamics_2022-1,goto_efficient_2023}.

The recursive approach has shown good performance in the previous studies~\cite{shillito_dynamics_2022-1,goto_efficient_2023}.
However, we find instances where the recursive approach gives the discontinuous dependence of physical quantities on photon number \(n\).
Consequently, we propose an approach that estimates the continuous dependence of physical quantities even in the case where the recursive approach does not.

The proposed method also determines the labeled state recursively.
From the eigenenergy \(\varepsilon_{p, n-1}\), one estimates a candidate for eigenenergy of \(\ket{\overline{p}, \overline{n}}\) as
\begin{align}
    \varepsilon^\prime_{p, n} = \varepsilon_{p, n-1} + (\varepsilon_{p, n-1} - \varepsilon_{p, n-2})
\end{align}
for \(n \geq 2\) and
\begin{align}
    \varepsilon^\prime_{p, 1} = \varepsilon_{p, 0} + \hbar \omega_c
\end{align}
for \(n = 1\).
The quantity \(\varepsilon_{p, n-1} - \varepsilon_{p, n-2}\) corresponds to the effective energy shift induced by adding a single photon.
Next, one selects every eigenstate whose eigenenergy \(\varepsilon \) satisfies \(|\varepsilon - \varepsilon^\prime_{p, n}| \leq \delta/2 \), where \(\delta \) is the energy window.
The parameter \(\delta \) should be large enough to detect sharp peaks coming from resonances and small enough not to
include unnecessary states.
If no eigenstate is found within the energy window, one selects the two eigenstates closest to the candidate.
For the selected eigenstates, one evaluates the expectation values of \(\hat{N}_q\).
The next labeled state \(\ket{\overline{p}, \overline{n}}\) is determined as the state with the closest expectation value to \(\braket{\overline{p}, \overline{n-1}|\hat{N}_q|\overline{p}, \overline{n-1}}\).
In determining \(\ket{\bar{p}, \bar{1}}\), the overlap approach is also available.
In short, the proposed method selects the state that makes the change of qubit occupancy as small as possible.

\section{Comparison of labeling methods\label{sec:comparison}}
In this section, we compare three labeling methods presented in Sect.~\ref{sec:method}, i.e., the overlap, the recursive~\cite{shillito_dynamics_2022-1}, and the proposed methods.
In the comparison, we choose the transmon as the qubit whose Hamiltonian is given by~\cite{bouchiat_quantum_1998,koch_charge-insensitive_2007,blais_circuit_2021,vool_introduction_2017}
\begin{align}
    \hat{H}_q = &4E_C \hat{N}^2_t- \frac{E_J}{2}\sum^\infty_{n=-\infty}\left( \ket{n}_t\bra{n+1}_t + \ket{n+1}_t \bra{n}_t\right).
\end{align}
Here, \(E_C\) is the charging energy, \(\hat{N}_t\) is the transmon charge operator, \(\ket{n}_t\) denotes the charge basis for the transmon, and \(E_J\) is the Josephson energy.
The transmon charge operator is given by
\begin{align}
    \hat{N}_t = \sum^\infty_{n=-\infty} (n - N_g) \ket{n}_t \bra{n}_t,
    \label{eq:Ntr}
\end{align}
where \(N_g\) is the offset charge.
By diagonalizing \(\hat{H}_q\) (with a truncated finite basis), one can obtain the eigenstates \(\ket{i}_q\) and the eigenvalues \(\hbar \omega_i\) of the qubit.
The transmon and the cavity are assumed to be capacitively coupled, i.e.,
\begin{align}
    \hat{H}_\mathrm{int} = i \hbar g \hat{N}_t (\hat{c}^\dagger - \hat{c}).
\end{align}
Here, \(g\) is the coupling between the transmon and the cavity.
For numerical computations, one has to truncate an infinite basis to some finite basis.
Hereafter, we consider states from \(\ket{-10}_t\) to \(\ket{10}_t\) for the charge basis of the transmon and the Fock state of the cavity up to \(\ket{350}_c\) unless otherwise specified.

\begin{figure}
\centering
\includegraphics[width=0.95\linewidth]{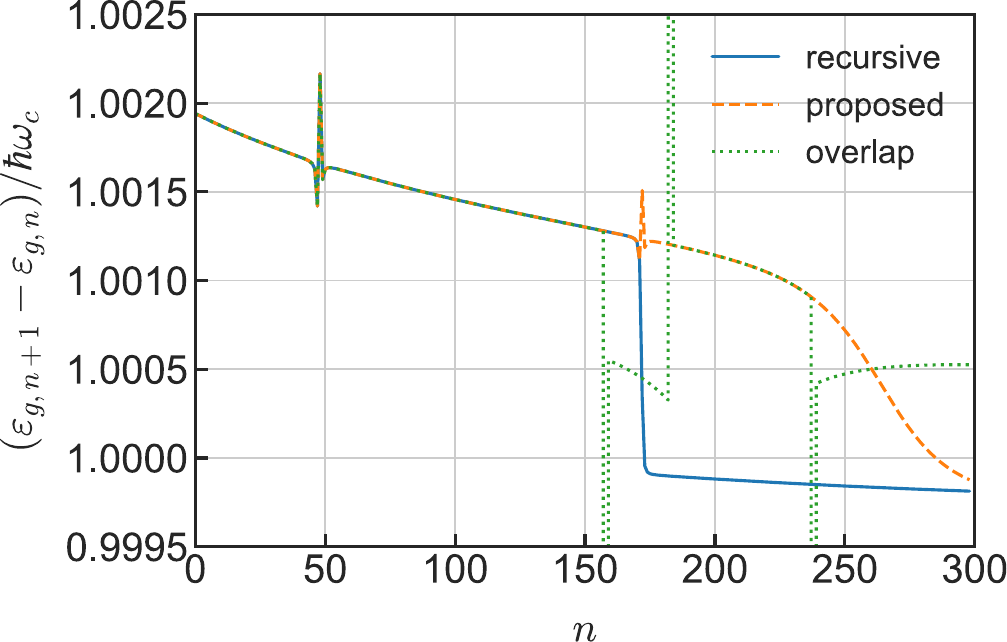}
\caption{(Color online) Photon-number dependence of the cavity frequencies in the transmon-cavity system obtained by the three labeling methods. The recursive calculations start from \(\ket{\overline{g}, \overline{0}}\). For the system parameters, we use \((E_C/\hbar \omega_c, E_J/\hbar \omega_c, g/\omega_c, N_g) = (5.0 \times 10^{-2}, 1.6, 2.5 \times 10^{-2}, 0.0)\).\ The energy window \(\delta \) is set to \(1.0 \times 10^{-2}\hbar \omega_c\) for the proposed method.\label{fig:comp_w_first}}
\end{figure}

Figure~\ref{fig:comp_w_first} gives the photon-number dependence of the cavity frequencies obtained by the three labeling methods.
The recursive calculations start from \(\ket{\overline{g}, \overline{0}}\).
We use \((E_C/\hbar \omega_c, E_J/\hbar \omega_c, g/\omega_c, N_g) = (5.0 \times 10^{-2}, 1.6, 2.5 \times 10^{-2}, 0.0)\) for the system parameters.
With the parameters, the occupations  of \(\ket{10}_t\) and \(\ket{-10}_t\) in the qubit eigenstate \(\ket{10}_q\) are only on the order of \(10^{-10}\). The qubit eigenstates energetically lower than \(\ket{10}_q\) are little affected by the truncation of the charge basis.
The energy window is set to \(\delta = 1.0 \times 10^{-2} \hbar \omega_c\).
The three labeling methods result in the same labeling up to \(n \sim 150\).
The sharp peak around \(n \sim 50\) comes from the resonance with the higher excited qubit states.
The transition to higher excited state occurs when the cavity photon number stays near the resonant peak.
However, such transition can be avoided by quickly passing the resonant point~\cite{shillito_dynamics_2022-1,goto_efficient_2023}.
Around \(n \sim 150\), the frequency given by the overlap method gives a discontinuous jump.
After the discontinuous jump, the cavity frequency obtained by the overlap method shows non-smooth dependence.
The observation of unphysical non-smooth behavior means that the overlap method fails to trace the labeled states \(\ket{\overline{g}, \overline{n}}\), and the failure implies that the labeled states \(\ket{\overline{g}, \overline{n}}\) are not close to \(\ket{g}_q \otimes \ket{n}_c\) for \(n \gtrsim 150\).

The other two methods give different cavity frequencies after \(n \sim 180\).
The cavity frequency given by the recursive method shows a sudden drop around \(n \sim 180\), while that estimated by the proposed method indicates the presence of the resonance at the point.
This observation suggests that the recursive method follows a resonant transition, while the proposed method passes through the transition.
Next, we try to confirm that the proposed method can actually avoid the resonant transitions by comparing the qubit occupation numbers \(\braket{\hat{N}_q}\) of the labeled states and that evolved under cavity driving.

To obtain the evolution of the qubit occupation number, we numerically simulate the Hamiltonian dynamics
\begin{align}
    \frac{d}{dt}\ket{\psi(t)} = -\frac{i}{\hbar} \hat{H}_\mathrm{dyn}(t) \ket{\psi(t)}
\end{align}
under the Hamiltonian 
\begin{align}
    \hat{H}_\mathrm{dyn}(t) = \hat{H} + \hbar E(e^{-i\omega_d t}\hat{c}^\dagger + e^{i\omega_d t}\hat{c})
\end{align}
with the initial state \(\ket{\psi(0)} = \ket{\overline{g}, \overline{0}}\).
Here, \(E\) is the amplitude of the driving field, and \(\omega_d\) is the drive frequency.
To reduce the number of the Fock states of the cavity used in numerical simulations, we apply the time-dependent unitary transformation
\begin{align}
    \hat{U}(t) = \exp\left[-\alpha(t)\hat{c}^\dagger + \alpha^*(t)\hat{c}\right]
\end{align}
to \(\ket{\psi(t)}\).
The displacement parameter \(\alpha(t)\) is determined by the differential equation
\begin{align}
    \frac{d \alpha(t)}{dt} = -i \left [\omega_c(\alpha(t) + \braket{\hat{c}(t)}_U) + ig \braket{\hat{N}_\mathrm{tr}(t)}_U + Ee^{-i\omega_d t}\right ],
\end{align}
where \(\braket{\hat{A}(t)}_U\) stands for \(\braket{\psi(t)|\hat{U}^\dagger(t)\hat{A}\hat{U}(t)|\psi(t)}\)~\cite{goto_efficient_2023}.
For the numerical simulation of the dynamics in this section, we use the Fock state of the cavity up to \(\ket{150}_c\), and the drive frequency is set to \(\omega_d / \omega_c = 1.0015\).

\begin{figure}
\centering
\includegraphics[width=0.95\linewidth]{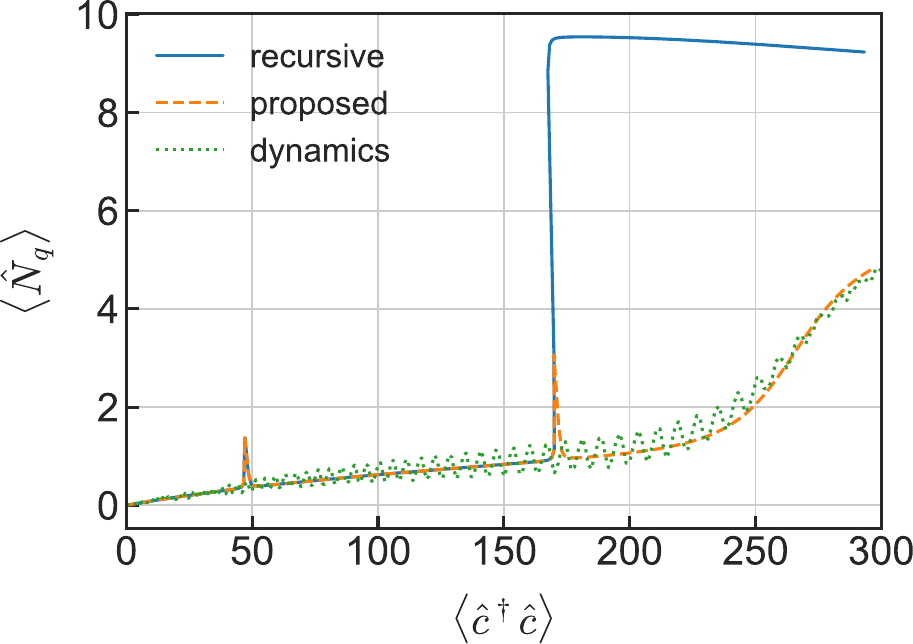}
\caption{(Color online) Photon-number dependence of the qubit occupation number in the transmon-cavity system obtained by the two labeling methods (recursive and proposed) and the evolution under cavity driving. The recursive calculations start from \(\ket{\overline{g}, \overline{0}}\). An initial state for the dynamics is also set to \(\ket{\overline{g}, \overline{0}}\). The parameters of the system are the same as those in Fig.~\ref{fig:comp_w_first}. The amplitude of the driving field \(E\) is set to \(5.0 \times 10^{-3}\omega_c\), and the drive frequency \(\omega_d\) is set to \(1.0015 \omega_c\).\label{fig:comp_Nq_first}}
\end{figure}

Figure~\ref{fig:comp_Nq_first} gives the photon-number dependence of the qubit occupation number obtained by the recursive and the proposed methods.
We also plot the trajectory in the \(\braket{\hat{N}_q}\)-\(\braket{\hat{c}^\dagger \hat{c}}\) plane under the cavity driving with \(E/\omega_c = 5.0 \times 10^{-3}\)in Fig.~\ref{fig:comp_Nq_first}.
The amplitude of the drive field \(E\) is determined strong enough not to be affected by resonances from higher excited states.
The qubit occupation numbers obtained from the two labeling methods show different behaviors for \(n \gtrsim 180\), and their behaviors are very similar to those in the cavity frequency shown in Fig.~\ref{fig:comp_w_first}.
The recursive method gives discontinuous increase of the qubit occupation number at \(n \sim 180\).
On the contrary, that obtained by the proposed method exhibits a sharp peak at the point.

\begin{figure}
\centering
\includegraphics[width=0.95\linewidth]{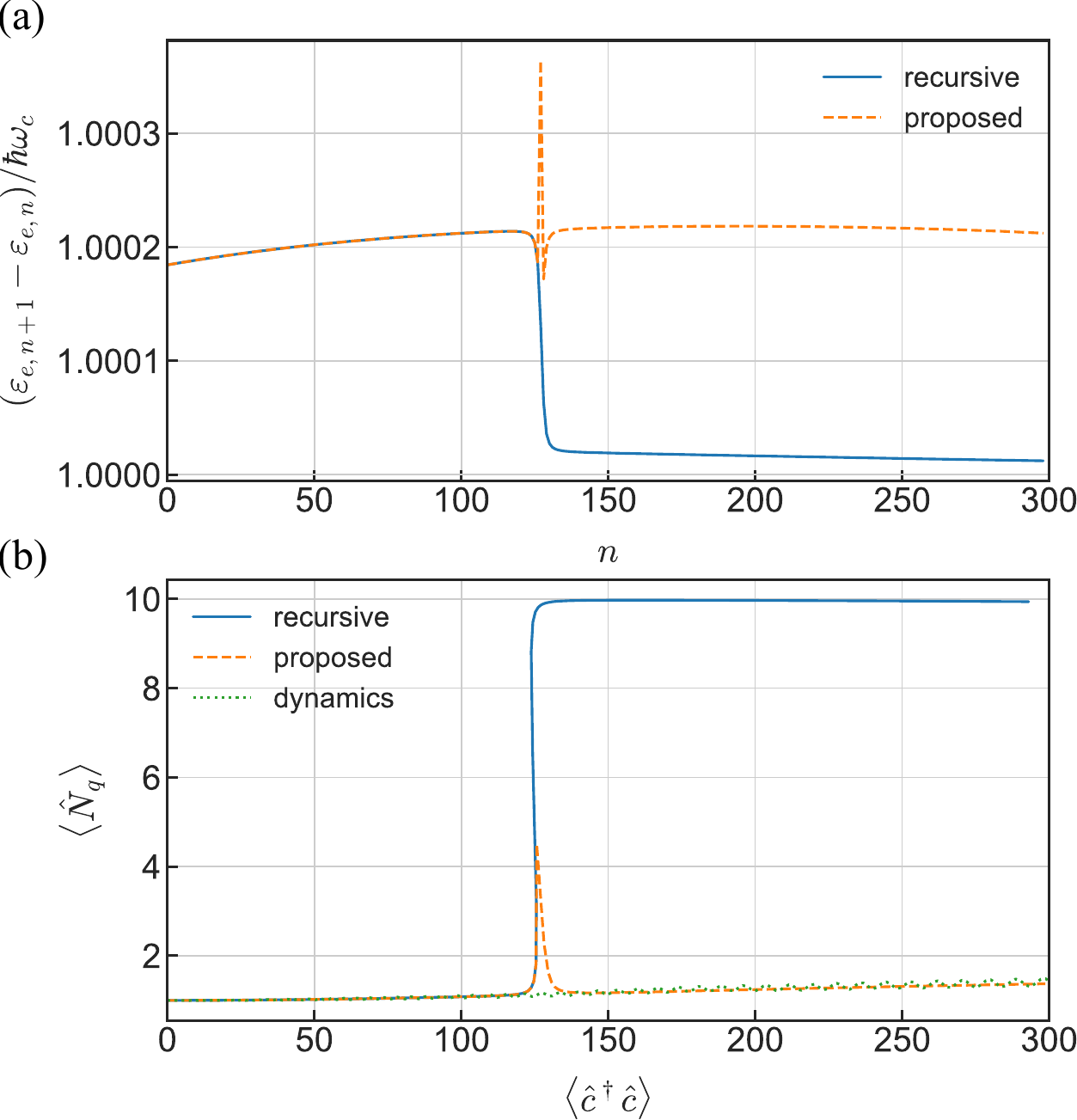}
\caption{(Color online) Photon-number dependence of (a) the cavity frequency and (b) the qubit occupation number in the transmon-cavity system obtained by the two labeling methods. For the qubit occupation number, the evolution under cavity driving is also plotted. The recursive calculations start from \(\ket{\overline{e}, \overline{0}}\). An initial state for the dynamics is also set to \(\ket{\overline{e}, \overline{0}}\). For the calculation, we use \((E_C/\hbar \omega_c, E_J/\hbar \omega_c, g/\omega_c, N_g, \delta/\hbar\omega_c, E/\omega_c, \omega_d / \omega_c) = (5.0 \times 10^{-2}, 2.0, 1.0 \times 10^{-2}, 0.1, 1.0 \times 10^{-2}, 5.0 \times 10^{-3}, 1.0002)\).\label{fig:comp_second}}
\end{figure}

Under cavity driving, the evolution of the qubit occupation number roughly follows the photon-number dependence obtained by the proposed method apart from the resonant peaks.
Therefore, we conclude that the proposed method can pass through the resonant transition and be applied to a broader situation compared to the recursive method.
We present another numerical example with parameters \((E_C/\hbar \omega_c, E_J/\hbar \omega_c, g/\omega_c, N_g, \delta/\hbar\omega_c, E/\omega_c, \omega_d / \omega_c) = (5.0 \times 10^{-2}, 2.0, 1.0 \times 10^{-2}, 0.1, 1.0 \times 10^{-2}, 5.0 \times 10^{-3}, 1.0002)\) starting from the excited state \(\ket{\bar{e}, \bar{0}}\).\ For this case with a finite charge offset, the same conclusion is obtained as summarized in Fig.~\ref{fig:comp_second}.
The recursive method follows a resonant transition, while the proposed method avoids the transition.
The proposed method can access the effective cavity frequency when the system passes through the resonant transition by sufficiently strong driving.

\section{energy-window dependence\label{sec:energy_window}}
\begin{figure}
\centering
\includegraphics[width=0.95\linewidth]{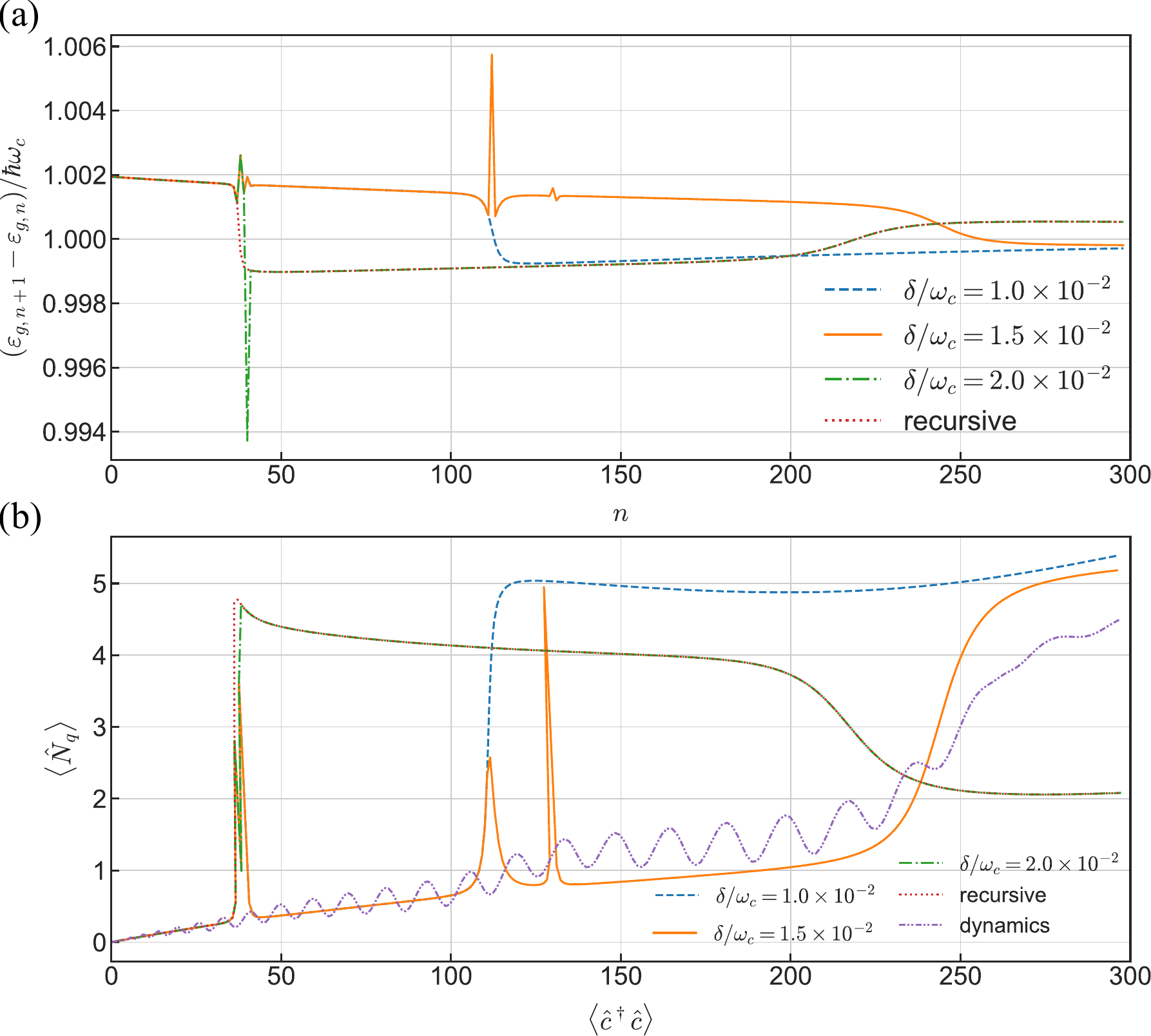}
\caption{(Color online) Photon-number dependence of (a) the cavity frequency and (b) the qubit occupation number in the transmon-cavity system obtained by the proposed labeling methods with different energy windows \(\delta \). For the qubit occupation number, the evolution under cavity driving is also plotted. The results obtained by the recursive labeling method are plotted for comparison. The labeling method starts from \(\ket{\overline{g}, \overline{0}}\). An initial state for the dynamics is also set to \(\ket{\overline{g}, \overline{0}}\).  For the calculation, we use \((E_C/\hbar \omega_c, E_J/\hbar \omega_c, g/\omega_c, N_g, E/\omega_c, \omega_d / \omega_c) = (5.0 \times 10^{-2}, 1.6, 2.5 \times 10^{-2}, 0.1, 1.0 \times 10^{-2}, 1.001)\).\label{fig:delta_dep}}
\end{figure}
Compared to the recursive approach, our proposed method introduces an additional parameter, the energy window \(\delta \).
This parameter should be large enough to include the eigenenergy affected by a higher excited state within the energy window and small enough to exclude eigenstates whose qubit occupancies are accidentally close to the target value.
The additional parameter can adjust the behavior of the labeling method, and thus the proposed method acquires adaptability.
In this section, we discuss how the energy window \(\delta \) affects the labeling.

To discuss the effects of the energy window, we observe the photon-number dependence of the effective cavity frequency and the qubit occupancy given by the proposed labeling method with different energy windows, as shown in Fig.~\ref{fig:delta_dep}.
We use the system parameters \((E_C/\hbar \omega_c, E_J/\hbar \omega_c, g/\omega_c, N_g) = (5.0 \times 10^{-2}, 1.6, 2.5 \times 10^{-2}, 0.1)\) and three distinct energy windows \(1.0 \times 10^{-2}\hbar \omega_c\), \(1.5 \times 10^{-2}\hbar \omega_c\), and \(2.0 \times 10^{-2}\hbar \omega_c\).
For the largest energy window \(2.0 \times 10^{-2} \hbar \omega_c\), the effective cavity frequency shows a discontinuous jump, and this jump suggests that the selected energy window is so large that a highly excited qubit state is included in the energy window.
The labeling with this energy window is very close to the labeling by the recursive method.
The recursive method can detect the resonance around \(n \sim 40\), while it cannot detect those around \(n \sim 110\) and \(n \sim 130\).
For the other two energy windows, a discontinuous behavior is not visible in the cavity frequency and the qubit occupancy except for resonant peaks.

By recalling the results presented in Sect.~\ref{sec:comparison}, a resonant peak of the cavity frequency is replaced with a sudden drop in a labeling following the resonant transition.
Based on this empirical rule, the labeling with the energy window \(\delta  = 1.5 \times 10^{-2}\hbar \omega_c\) should be appropriate to evaluate the effective cavity frequency with avoiding resonant peaks.
This expectation can be verified by comparing the trajectory obtained by the dynamics under cavity driving.
The red-dotted line in Fig.~\ref{fig:delta_dep}(b) is the trajectory in the \(\braket{\hat{N}_q}\)-\(\braket{\hat{c}^\dagger \hat{c}}\) plane under the cavity driving with \(E/\omega_c = 1.0 \times 10^{-2}\) and \(\omega_d/\omega_c = 1.001\).
We use the Fock state of the cavity up to \(\ket{500}_c\) for the numerical simulation of the dynamics.
The trajectory obtained by the dynamics follows the photon number dependence of the qubit occupancy with the energy window \(\delta/\hbar \omega_c = 1.5 \times 10^{-2}\).
The dynamics ensure that the labeling with \(\delta/\hbar \omega_c = 1.5 \times 10^{-2}\) is appropriate to avoid resonant transitions.
This rough matching with the dynamics also implies that a drop of the cavity frequency around \(n \sim 240\) will not be replaced with a resonant peak.

When two labelings with energy windows \(\delta_1\) and \(\delta_2\) show different behaviors at a resonant point, the level repulsion between resonant states can be bounded by using the parameters \(\delta_1 \) and \(\delta_2\).
The two labelings with energy windows \(\delta_1\) and \(\delta_2\) are supposed to give the same label up to the photon number \(n-1\).
Then, the next eigenenergies \(\varepsilon_{p, n}\) are assumed to be different in the two labelings, and we represent them as \(E_r +\)(\(-\))\(\Delta / 2\) for \(\delta_1\)(\(\delta_2\)).
Here, \(E_r\) is the resonant energy and \(\Delta \) is the level repulsion emerged from the interaction between resonant states.
We choose the energy window for the larger eigenenergy as \(\delta_1\).
Consequently, the level repulsion is positive, \(\Delta > 0\).
In such situations, the energy window conditions \(|\varepsilon_{p,n} - \varepsilon^\prime_{p,n}| \leq \delta / 2\) for the two labelings can be rewritten as
\begin{align}
    2(\varepsilon^\prime_{p,n} - E_r) - \delta_1 &\leq \Delta \leq 2(\varepsilon^\prime_{p,n} - E_r) + \delta_1\label{eq:ineq_1}\\
    -2(\varepsilon^\prime_{p,n} - E_r) - \delta_2 &\leq \Delta \leq -2(\varepsilon^\prime_{p,n} - E_r) + \delta_2\label{eq:ineq_2}.
\end{align}
We note that a candidate for eigenenergy \(\varepsilon^\prime_{p, n}\) is the same for the two labelings up to \(n\).
Since \(\Delta \) is positive, the inequality \(-2(\varepsilon^\prime_{p,n}-E_r) < \delta_1\) is satisfied from Eq.~\eqref{eq:ineq_1}.
With the inequality and Eq.~\eqref{eq:ineq_2}, one can bound the level repulsion \(\Delta \) as
\begin{align}
    0 < \Delta < \delta_1 + \delta_2.
\end{align}
Especially when the parameters \(\delta_1\) and \(\delta_2\) have the same order of magnitude \(\delta_1 \sim \delta_2 \sim \delta\), the bound for the repulsion is given as
\begin{align}
    0 < \Delta \lesssim 2\delta.
\end{align}
Thus, the energy window parameter \(\delta \) roughly bounds the allowable size of level repulsions that the proposed labeling method can follow.
To change labeling behaviors at a resonant point, the parameter \(\delta \) should be varied by the same magnitude as \(\Delta \).

Through the example presented in this section, one can find the importance of using some energy windows and comparing those results.
When a sudden change in the cavity frequency is replaced with a peak in another labeling, the labeling with the sudden change would follow a resonant transition. 
This empirical rule is useful for choosing an appropriate energy window to avoid resonant transitions without accessing dynamics.
For the numerically expensive but definite confirmation, one can compare the results of labeling to those obtained from dynamics.

\section{Offset-charge dependence of resonance\label{sec:result}}
As we have already observed, the indication of the resonance to higher excited states can be seen in the photon-number dependence of the cavity frequency obtained by the labeling methods.
Such resonance can lead to leakage from the computational space~\cite{sank_measurement-induced_2016,khezri_measurement-induced_2023,shillito_dynamics_2022-1}.
The identification of its location would be an essential in designing the superconducting qubits.
Since the labeling method proposed in this study can be applied to a broader situation compared to the previous method~\cite{shillito_dynamics_2022-1}, the proposed method is appropriate for the identification.

In this section, we concentrate our attention on the offset-charge \(N_g\) dependence of the resonant points.
The offset charge contains contributions from charge impurities.
One of the important characteristics of the transmon superconducting qubit is robustness of its resonant frequency to the offset-charge noise.
In other words, the energy difference between the first excited and the ground states has little dependence on the offset charge in the transmon regime \(E_J / E_C \gg 1\).
It should be noted that the eigenenergies of higher excited states depend on the offset charge~\cite{blais_circuit_2021}.
The presence of the chaotic layer in the spectrum of a driven transmon introduces the offset-charge dependence to the dynamics of low-lying qubit states indeed~\cite{cohen_reminiscence_2023}.
Consequently, the locations of resonances to the higher excited states are expected to depend on the offset-charge noise.

Because of the infinite sum in the charge operator defined in Eq.~\eqref{eq:Ntr}, the operators before and after the shift of the offset charge by unity, \(N_g \to N_g +1\), are unitarily equivalent, i.e.,
\begin{align}
    \begin{aligned}
    \hat{V}^\dagger_\mathrm{shift} \left \{\sum^\infty_{n=-\infty} [n - (N_g +1 )] \ket{n}_t \bra{n}_t \right \} \hat{V}_\mathrm{shift}\\
     = \sum^\infty_{n=-\infty} (n - N_g) \ket{n}_t \bra{n}_t
    \end{aligned}
\end{align}
with the unitary transformation \(\hat{V}_\mathrm{shift} = \sum \ket{n}_t \bra{n-1}_t\).
The unitary transformation \(\hat{V}_\mathrm{shift}\) does not change the Josephson energy term \((E_J/2)\sum (\ket{n}_t \bra{n+1}_t + \ket{n+1}_t \bra{n}_t)\).
Thus, the spectrum of the transmon-cavity system is invariant under the shift.
Besides, the infiniteness of the summation leads to another symmetry of the charge operator: The operator with the sign inverted offset charge is unitary equivalent to the original operator with the negative sign, i.e.,
\begin{align}
    \begin{aligned}
    \hat{V}^\dagger_\mathrm{invert} \left \{\sum^\infty_{n=-\infty} (n + N_g) \ket{n}_t \bra{n}_t \right \} \hat{V}_\mathrm{invert}\\
     = -\sum^\infty_{n=-\infty} (n - N_g) \ket{n}_t \bra{n}_t
    \end{aligned}
\end{align}
with the unitary transformation \(\hat{V}_\mathrm{invert} = \sum \ket{n}_t \bra{-n}_t\).
The unitary transformation \(\hat{V}_\mathrm{invert}\) does not change the Josephson energy term either.
Since the negative sign can be absorbed into the definition of the cavity annihilation and creation operators, the spectrum of the transmon-cavity system is invariant under the sign inversion of the offset charge.
From the periodicity and the symmetry, it is enough to investigate the region \(0 \leq N_g \leq 0.5\) for the identification.

\begin{figure}
\centering
\includegraphics[width=0.98\linewidth]{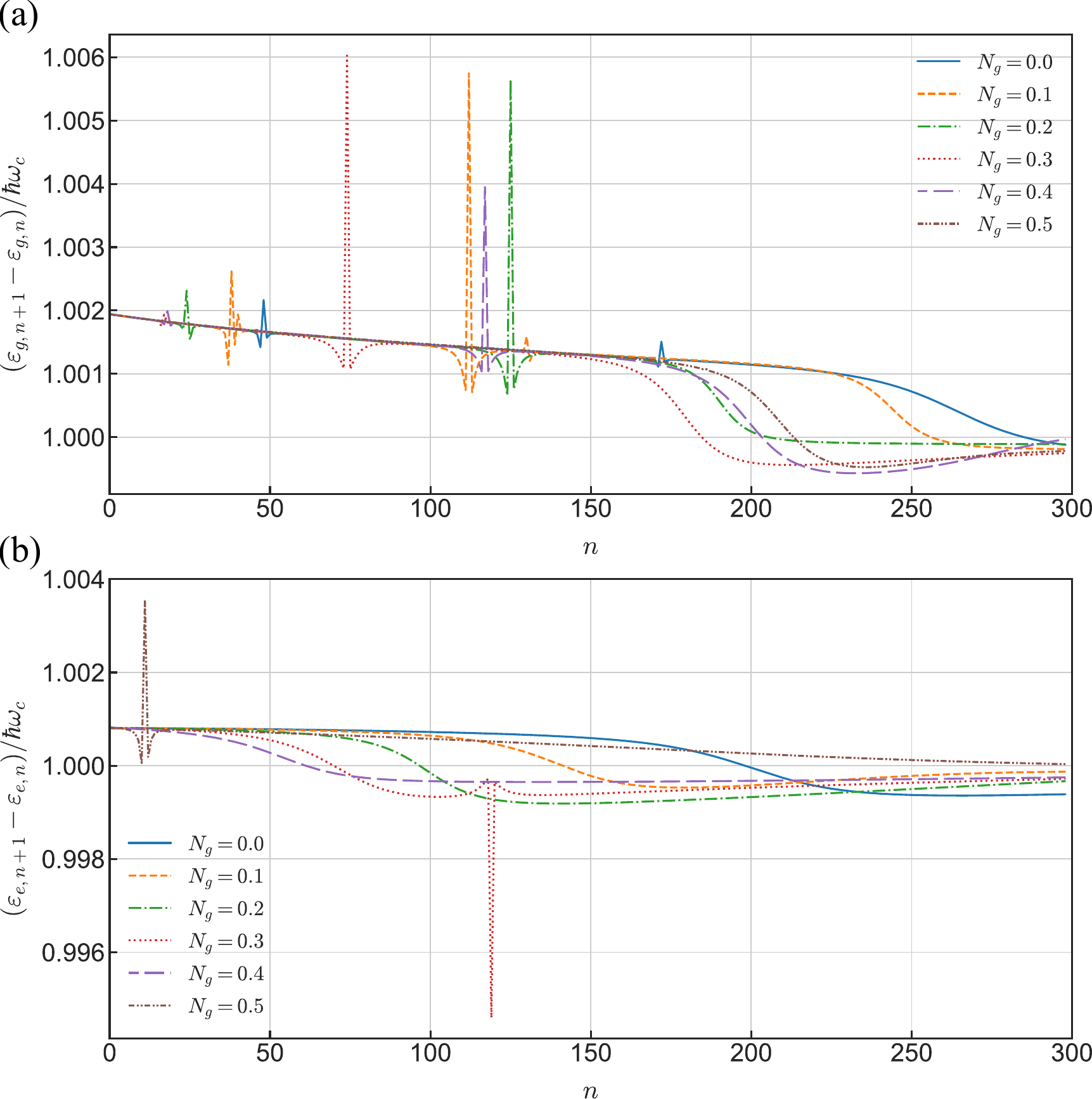}
\caption{(Color online) Photon-number dependence of the cavity frequency with the offset charges from \(0\) to \(0.5\) for (a) the ground-state ladder \(\ket{\overline{g}, \overline{n}}\) and (b) the excited-state ladder \(\ket{\overline{e}, \overline{n}}\). We use \((E_C/\hbar \omega_c, E_J/\hbar \omega_c, g/\omega_c) = (5.0 \times 10^{-2}, 1.6, 2.5 \times 10^{-2})\) for the system parameters. For the cases \(N_g = 0.1\) in the ground-state ladder and \(N_g = 0.3\) in the excited-state ladder, the energy window \(\delta \) is set to \(1.5 \times 10^{-2} \hbar \omega_c\). For the other cases, we set \(\delta \) to \(1.0 \times 10^{-2} \hbar \omega_c\).\label{fig:NgDep}}
\end{figure}

Figure~\ref{fig:NgDep} shows the photon-number dependence of the cavity frequency with the offset charges from 0 to 0.5 for the ground- (excited-)state ladder \(\ket{\overline{g}, \overline{n}}\) (\(\ket{\overline{e}, \overline{n}}\)).
For the ground-state ladder \(\ket{\overline{g}, \overline{n}}\), the cavity frequencies for the different offset charges do not show visible difference up to \(n \sim 150\) apart from the resonant peaks.
The little dependence on the offset charge implies that the weights of higher transmon excited states in the labeled states \(\ket{\overline{g}, \overline{n}}\) are small for \(n \lesssim 150\) (apart from the resonant point).
The location of the resonant point strongly depends on the offset charge, as expected.
The resonant point can fluctuate between \(n \sim 15\) and \(n \sim 130\) without controlling the offset charge. 
Therefore, a transition to a higher excited state may take place with a relatively small photon number \(n \sim O(10)\).

For the excited-state ladder \(\ket{\overline{e}, \overline{n}}\), the cavity frequencies show visible difference for \(n \gtrsim 20\).
This implies that the weights of higher transmon excited states are not negligible in this region.
For the case with \(N_g = 0.5\), the resonant peak is located around \(n \sim 10\).
Consequently, a transition to a higher excited state may take place with a relatively small photon number also in the case of the labeled states \(\ket{\overline{e}, \overline{n}}\).
In this way, the photon-number dependence of the cavity frequency obtained by the labeling method provides the information on photon number where the transitions leading to leakage from the computational space can occur.

\section{Summary\label{sec:summary}}
In this paper, we proposed a new method to label the eigenstates of a qubit-cavity system based on the continuity of qubit occupancy.
Starting from the eigenstate without photon occupation, the system is expected to follow the labeled states under adiabatic injection of photons into the cavity.
Besides, the effective resonant frequency of the cavity with photon occupation can be estimated from the eigenenergies of the labeled states.
Our proposed method finds candidates for the next labeled state based on their eigenenergies.
Then, one selects the next labeled state that makes the change of the qubit occupancy as small as possible.
Comparing the obtained photon-number dependence of the cavity frequency and the evolution under cavity driving, we confirmed that our proposed scheme can be applied to a broader situation compared to the previous approach~\cite{shillito_dynamics_2022-1} in the sense that our proposed scheme can access a case where a system passes through resonant transitions that the previous approach cannot avoid.

With the proposed method, we investigated the offset-charge dependence of the cavity frequency in the transmon-cavity system.
In the system we studied, the location of the resonance fluctuates depending on the offset charge.
Even though the qubit resonant frequency of the transmon is robust to the charge noises, the resonance to higher excited states is not.
Without controlling the offset charge, it is possible to encounter a resonance to higher excited states with a relatively small photon number \(n \sim 10\).

The photon-number dependence of the cavity frequency and the location of the resonance to higher excited states are essential information for designing superconducting qubits.
Our proposed method is applicable to other superconducting qubits such as the fluxonium~\cite{manucharyan_fluxonium_2009} or qubits with \(d\)--wave superconductors~\cite{patel_d-mon_2024,brosco_superconducting_2024}.
Their robustness to the charge noise, including effects from higher excited states, can be evaluated with a labeling method.
A stable labeling method is essential in assessing the robustness of existing or new high-performance superconducting qubits.

\begin{acknowledgments}
    We thank K. Sakai for the fruitful discussions.
    This work was financially supported by JST Moonshot R\&D Grant Numbers JPMJMS2061 and JPMJMS2067.
\end{acknowledgments}
%

\end{document}